\documentclass[lettersize,journal]{IEEEtran}
\usepackage{amsmath,amsfonts}
\usepackage{algorithmic}
\usepackage{algorithm}
\usepackage{array}
\usepackage[caption=false,font=normalsize,labelfont=sf,textfont=sf]{subfig}
\usepackage{textcomp}
\usepackage{stfloats}
\usepackage{url}
\usepackage{verbatim}
\usepackage{graphicx}
\usepackage{cite}
\usepackage{tabularx}
\newcolumntype{L}{>{\raggedright\arraybackslash}X}
\usepackage{graphicx}
\usepackage[export]{adjustbox}
\usepackage{flushend}
\usepackage[acronym,nomain,nonumberlist]{glossaries}
\usepackage{fancyhdr}

\makeglossaries

\newacronym{mn}{MN}{master node}
\newacronym{sn}{SN}{secondary node}
\newacronym{ntn}{NTN}{non-terrestrial network}
\newacronym{tn}{TN}{terrestrial network}
\newacronym{nr}{NR}{new radio}
\newacronym{mc}{MC}{multi-connectivity}
\newacronym{itu}{ITU}{the International Telecommunications Union}
\newacronym{ngso}{NGSO}{non-geosynchronous orbit}
\newacronym{gso}{GSO}{geosynchronous orbit}
\newacronym{meo}{MEO}{medium Earth orbit}
\newacronym{haps}{HAPS}{high altitude platform station}
\newacronym{vsat}{VSAT}{very small aperture terminal}
\newacronym{leo}{LEO}{low Earth orbit}
\newacronym{phy}{PHY}{physical}
\newacronym{mac}{MAC}{medium access control}
\newacronym{rlc}{RLC}{radio link control}
\newacronym{pdcp}{PDCP}{packet data convergence protocol}
\newacronym{mrdc}{\hbox{MR-DC}}{multi-radio dual connectivity}
\newacronym{nrdc}{\hbox{NR-DC}}{new radio dual connectivity}
\newacronym{cn}{CN}{core network}
\newacronym{bs}{BS}{base station}
\newacronym{mmtc}{mMTC}{massive machine-type com\-mu\-ni\-ca\-tions}
\newacronym{urllc}{URLLC}{ultra-reliable low latency communications}
\newacronym{embb}{eMBB}{enhanced mobile broadband}
\newacronym{hrcs}{\hbox{HRC-s}}{high reliability communications via satellite}
\newacronym{embbs}{\hbox{eMBB-s}}{enhanced mobile broadband via satellite}
\newacronym{mmtcs}{\hbox{mMTC-s}}{massive machine-type communications via satellite}
\newacronym{trp}{TRP}{transmission and reception point}
\newacronym{ca}{CA}{carrier aggregation}
\newacronym{atsss}{ATSSS}{access traffic steering, switching, and splitting}
\newacronym{daps}{DAPS}{dual active protocol stack}
\newacronym{3gpp}{3GPP}{the 3rd Generation Partnership Project}
\newacronym{ai}{AI}{artificial intelligence}
\newacronym{ml}{ML}{machine learning}
\newacronym{eutra}{E-UTRA}{evolved universal terrestrial radio acces}
\newacronym{dc}{DC}{dual connectivity}
\newacronym{ue}{UE}{user equipment}
\newacronym{gnb}{gNB}{next-generation node B}
\newacronym{ts}{TS}{technical specification}
\newacronym{tr}{TR}{technical report}
\newacronym{ho}{HO}{handover}
\newacronym{ra}{RA}{random access}
\newacronym{arq}{ARQ}{automatic repeat request}
\newacronym{harq}{HARQ}{hybrid automatic repeat request}
\newacronym{sdap}{SDAP}{service data adaptation protocol}
\newacronym{5ga}{\hbox{5G-A}}{5G-advanced}
\newacronym{b5g}{B5G}{beyond 5G}

\newacronym{isl}{ISL}{inter-satellite link}
\newacronym{gnss}{GNSS}{global navigation satellite system}
\newacronym{pd}{PD}{packet duplication}

\newacronym{}{}{}

\usepackage[roman]{parnotes}
\makeatletter
\def\parnoteclear{%
    \gdef\PN@text{}%
    \parnotereset
}
\makeatother

\fancypagestyle{FirstPage}{

\lfoot{\small \copyright 2023 IEEE. Personal use of this material is permitted. Permission from IEEE must be obtained for all other uses, in any current or future media, including reprinting/republishing this material for advertising or promotional purposes, creating new collective works, for resale or redistribution to servers or lists, or reuse of any copyrighted component of this work in other works. DOI: 10.1109/MCOM.001.2300581}

}

\begin{document}

\title{Toward Multi-Connectivity in Beyond 5G Non-Terrestrial Networks: Challenges and Possible Solutions}

\author{Mikko Majamaa
\thanks{This work has been partially funded by the European Union Horizon-2020 Project DYNASAT (Dynamic Spectrum Sharing and Bandwidth-Efficient Techniques for High-Throughput MIMO Satellite Systems) under Grant Agreement 101004145. The views expressed are those of the authors and do not necessarily represent the project. The Commission is not liable for any use that may be made of any of the information contained therein.}
\thanks{\textit{The author is with Magister Solutions, and University of Jyväskylä, Finland.}}
}



\maketitle

\begin{abstract}
\Glspl{ntn} will complement \glspl{tn} in 5G and beyond, which can be attributed to recent deployment and standardization activities. Maximizing the efficiency of \gls{ntn} communications is critical to unlock its full potential and reap its numerous benefits. One method to make communications more efficient is by the usage of \gls{mc}, which allows a user to connect to multiple base stations simultaneously. It is standardized and widely used for \glspl{tn}, but for \gls{mc} to be used in the \gls{ntn} environment, several challenges must be overcome. In this article, challenges related to \gls{mc} in \glspl{ntn} are discussed, and solutions to the identified challenges are proposed.
\end{abstract}



\section*{Introduction}

\glsresetall

\thispagestyle{FirstPage}

Mobile communications and access to the Internet play a critical role in a prosperous society. At the individual level, this can be explained by the increased opportunities that access to the Internet brings, such as educational and employment opportunities. At the societal level, it can be explained by factors such as business transformation and economic modernization. Yet, according to the statistics published by \gls{itu}, there were 2.7 billion people in the world who had no access to the Internet in 2022 \cite{itu27}. To this end, \glspl{ntn} can be used to provide connectivity to un(der)served areas.

Recent standardization and deployment activities have made the deployment of 5G and beyond through \glspl{ntn} a practical possibility \cite{9914764}. The standardization efforts enhance cooperation between terrestrial and satellite operators. In the future, the different networks can be seen as complementary rather than competing, while converging into one indistinguishable network from the user's point of view. In addition, \gls{ngso} satellites, in particular, have been the subject of intense research and deployment interest in recent years, due to advances in launch vehicles that allow many small satellites to fit into a single launch, and the mass production of satellites, which lowers production costs.

\Glspl{ntn} offer a variety of ways to improve the performance of wireless communications. Extending coverage to unserved and underserved areas means services such as backhaul connections to remote networks, direct-to-handheld services, maritime coverage, and remote IoT connectivity. In addition to extending coverage, \glspl{ntn} can provide load balancing to overloaded \glspl{tn}. In cases of natural disasters or emergencies, \glspl{ntn} can provide critical access when \glspl{tn} are out of coverage.

To ensure the efficiency of \gls{ntn} communications, research is needed to explore methods on how to achieve this. \Gls{mc} is a promising method that allows a \gls{ue} to connect to multiple \glspl{bs} simultaneously. For example, \gls{mc} in \glspl{ntn} can be used in load balancing  \cite{9143093}, and to enhance service continuity \cite{10199206} and energy efficiency  \cite{10106450}. Further, \gls{mc} for throughput enhancement may be critical to enable applications with high throughput requirements to cell edge users, such as live video streaming, where a single link cannot achieve the required performance \cite{10195478}. \Gls{mc} between satellite and cellular networks in rural areas can help reduce latency, increase availability, resiliency, and reliability, and thus help meet the low latency requirements of applications such as environmental and livestock monitoring \cite{10008752}.

\Gls{mc} can be implemented at different levels such as \gls{phy} layer, \gls{mac} layer, \gls{pdcp} layer, or \gls{cn} level. \Gls{pdcp} layer \gls{mc} solutions are attractive because they can quickly adapt to changing radio conditions. \Gls{mrdc} is one such solution and is the main focus of the article. In \gls{mrdc}, a \gls{ue} is connected to two \glspl{bs} one providing 5G access and the other 4G/5G access. \Gls{mc} in \glspl{ntn} can include only \gls{ntn} nodes or a mix of \gls{tn} and \gls{ntn} nodes.

\Gls{mrdc} has been standardized for \glspl{tn}, yet its standardization for \glspl{ntn} has not been actualized to date. Additionally, although there is a substantial body of literature addressing \gls{mc} in \glspl{tn} \cite{s22197591}, analogous comprehensive research exploring \gls{mc} in \glspl{ntn} remains comparatively underdeveloped. This article aims to help fill this research gap by providing a comprehensive introduction and discussion on \gls{mc} in \glspl{ntn}, including related challenges and possible solutions. To achieve this, the article examines 5G and beyond specifications, existing literature, and insights. Notably, this article is the first published tutorial/survey article related to \gls{mc} in \glspl{ntn}, to the best of the author's knowledge. Although the examples in this article mainly center around \gls{ngso}-based \glspl{ntn}, given their inherently more complex nature due to their high velocity relative to users, the discussion can be applied to other types of \glspl{ntn}.

The article's structure is as follows: first, 5G and beyond networks are introduced, followed by a discussion on the way toward \gls{mc} in \glspl{ntn}. Then, considerations about \gls{mc} in \glspl{ntn} are analyzed, including challenges and potential solutions. Finally, the article is concluded.

\section*{5G and Beyond Networks}

\Gls{3gpp} is a leading standards body for mobile communications standards. The purpose of standardization is to ensure high-quality systems and that the various network operators and equipment manufacturers can work together seamlessly. \Gls{3gpp} \mbox{Rel-15} marked the beginning of the 5G era in terms of \gls{3gpp} standardization. Subsequent releases 16 and 17 continued by enhancing the features. Releases 15 through 17 focused on 5G, while the ongoing \mbox{Rel-18} marks the beginning of \gls{5ga} and \gls{b5g}. For example, \mbox{Rel-18} addresses network energy savings, mobility, and coverage enhancements, \gls{ai}/\gls{ml} for next-generation radio access network, and dynamic spectrum sharing enhancements \cite{9927255}. In addition, future 6G use cases include holographic telepresence applications, tactile Internet, teleoperated driving, nanonetworks, and a surge in the number of IoT devices \cite{9040264}.

History was made in March 2022 when \gls{3gpp} included \glspl{ntn} in its specifications in its Release~17 (\mbox{Rel-17}), although it should be noted that \hbox{Rel-15} and \hbox{Rel-16} included necessary preparatory work in the form of \glspl{tr}. \Gls{ntn} work in \gls{3gpp} covers \gls{leo}, \gls{meo}, and \gls{gso} satellite communications, as well as airborne vehicles such as \glspl{haps}. \mbox{Rel-17} provides specifications for \gls{nr}-based satellite access in S- and L-bands, where \gls{nr} is the air interface of 5G. The focus is on transparent payload architecture and frequency division duplexing serving handheld terminals. \mbox{Rel-18} will further enhance \gls{nr} operations in \glspl{ntn}, for example, by improving coverage to Ka-bands serving \glspl{vsat}, as well as addressing mobility and service continuity.

\section*{Toward Multi-Connectivity in Non-Terrestrial Networks}

This section discusses the path toward \gls{mc} in \glspl{ntn} in the context of 5G technology. The goal of using \gls{mc} is to improve end-user service performance. Various forms of \gls{mc} have already been standardized for \glspl{tn} \cite{9665425}.

\Gls{mc} in 5G can be used on different protocol layers. Multi-\gls{trp} (\gls{ts}~38.300) is a \gls{phy} layer \gls{mc} solution. \Gls{ca} (\gls{ts}~38.331) can be used to achieve \gls{mc} on the \gls{mac} layer. \Gls{pdcp} layer \gls{mc} solutions include \gls{nrdc} (\gls{ts}~37.340), where the \gls{ue} is connected to two \glspl{bs} providing 5G access, \gls{mrdc} (\gls{ts}~37.340), where the \gls{ue} is connected to two \glspl{bs} providing 5G and 4G/5G access, and \gls{daps} (\gls{ts}~38.300), where the \gls{ue} is connected to two 5G \glspl{bs} to ensure a seamless handover. \Gls{mc} can also be achieved at the \gls{cn} in terms of \gls{atsss} (\gls{ts}~24.193), where simultaneous \gls{3gpp} and non-\gls{3gpp} access is provided. 

Each \gls{mc} solution has some limitations. The \gls{phy} layer solutions have strict latency and synchronization requirements. The \gls{mac} layer solutions introduce increased scheduler complexity. The \gls{pdcp} layer solutions require additional hardware and software capabilities. The \gls{cn}-based \gls{mc} can be problematic because the \gls{cn} may not be able to quickly adapt to dynamic radio link conditions.

A \gls{cn}-based \gls{mc} for \glspl{ntn}, that is, upper-layer traffic steer, switch, and split over dual \gls{3gpp} access (\gls{tr}~22.841), will be included in the upcoming \hbox{Rel-19}. This type of \gls{mc} is similar to \gls{atsss}. In addition, \gls{mrdc} supporting \gls{ntn} was also planned for \hbox{Rel-19}, but was scoped out. Although scoped out of \hbox{Rel-19}, it is a promising solution (and a candidate for future releases) because of its ability to quickly adapt to dynamic radio link conditions. In this article, the focus is primarily on the \gls{pdcp} layer \gls{mc}, namely, \gls{mrdc}, and the term \gls{mc} is used interchangeably to mean \gls{mrdc} in the article unless otherwise noted. The analysis of \gls{mc} implementations on different layers is left for future work.

\Gls{mrdc} is a generalization of \gls{eutra} \gls{dc}, that is, 4G \gls{dc}. In \gls{mrdc}, one of the nodes serves as a \gls{mn} and the other as a \gls{sn}. One of the nodes provides 5G access and the other either 5G or 4G access. The nodes are connected via the Xn interface for control signaling as well as to steer traffic from the \gls{mn} to the \gls{sn}. In principle, more than two connections to a user could be formed but hardware and software requirements can be a limiting factor. Further, according to TR~38.821, the \gls{mrdc} specification for \glspl{tn} (TS~37.340) may need to be adapted to support NTN MC, for example, to accommodate extended latency, variable latency (e.g. Xn interface traversing multiple satellites on different orbital planes), and differentiated delay between different nodes involved.

The \gls{sn} addition for a \gls{ue} is triggered by the \gls{mn} (i.e., the current serving node of the \gls{ue}) that sends \gls{sn} addition request to a candidate SN. After the \gls{sn} addition procedure is completed, the \gls{mn} can send data to the \gls{sn} through the Xn interface. The \gls{sn} then forwards the data down the protocol stack to allocate the necessary resources and finally send it over the air to the \gls{ue}. This does not significantly change the \gls{bs} architecture. On the \gls{ue} side, in the downlink direction, the \gls{ue} must be able to receive data from the two connections, which means that the \gls{ue} must have two receiving antenna elements. At the protocol stack level, this means that the \gls{ue} must receive the two different transmissions and then combine them at the \gls{pdcp} layer which has an impact on the \gls{ue}’s architecture. When uplink \gls{mc} is considered, the \gls{ue} must be able to transmit to the \gls{mn} and to the \gls{sn} simultaneously.

\begin{figure*}[!t]
\centering
\includegraphics[width=.8\linewidth, frame]{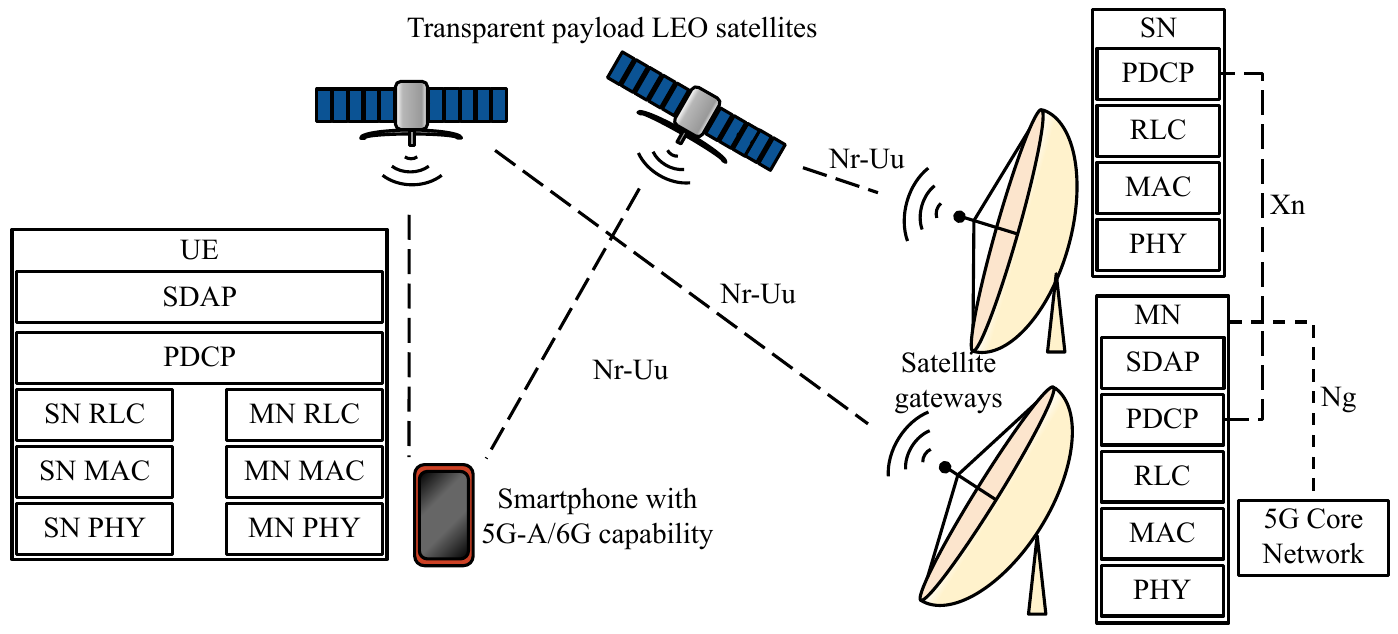}
\caption{\gls{mc} illustrated in the \gls{ntn} environment. The figure shows the related protocol stacks, where changes to the \gls{ue}’s architecture can be seen: the \gls{ue} must be able to receive transmissions from multiple sources, namely the \gls{mn} and \gls{sn} in the downlink direction, and similarly, to transmit to these different nodes when uplink \gls{mc} is considered. A transparent payload architecture is considered, so the service and feeder links use the Nr-Uu interface. The \gls{mn} and \gls{sn} are connected via the Xn interface for the exchange of control plane and user plane data. The \gls{mn} is connected to the \gls{cn} via the Ng interface. Adapted from \cite{10195478}.}

\label{illustration}
\end{figure*}

\Gls{mc} in \glspl{ntn} is illustrated in Fig.~\ref{illustration}. Some of the main functions of the layers in the protocol stack shown in the figure are briefly summarized as follows \cite{https://doi.org/10.1002/sat.1469}:
\begin{itemize}
\item{\Gls{sdap}, \gls{pdcp}, \gls{rlc}. Responsible for assembling and reordering packets, security, \gls{arq}, and integrity protection.}
\item{\gls{mac}. Defines how to access the medium. Responsible for logical channel prioritization, \gls{ra}, and retransmissions through \gls{harq}.}
\item{\gls{phy}. Physical layer operations, such as (de)modulation, (inverse) fast Fourier transform, and orthogonal frequency division multiplexing operations, that is, the responsibility for transmitting the raw data over the air.}
\end{itemize}

\section*{Considerations for Multi-Connectivity in Non-Terrestrial Networks}

\Gls{mc} in \glspl{tn} is well-studied and documented but few works have explored \gls{mc} in \gls{ntn}, and the unique challenges of this environment need to be considered. The problems and solutions for \gls{mc} in \gls{ntn} are discussed below and summarized in Table~\ref{table1}. Please note that the solutions listed in the table with specific references are derived from existing scientific work. Solutions without such references represent original contributions proposed by the author.

\begin{table*}[hbt!]
\caption{Problems and Solutions for \gls{mc} in \glspl{ntn}.}
\label{table1}
\begin{tabularx}{\linewidth}{|l|L|L|}
\hline
\textbf{Attribute} & \textbf{Problem} & \textbf{Solutions} \\ \hline

Different node types & \gls{mc} in \glspl{ntn} may include a heterogeneous set of nodes involved (transparent/regenerative payload satellites and terrestrial nodes). & \begin{itemize}
    \item{Choosing the possible combination so that it is feasible in terms of possibly required: 1) increased buffering; 2) formation of ISLs; and 3) increased path length for the data transmission.}
\end{itemize} \\ \hline

Delay differences & Delay may vary between the different paths. & 
\begin{itemize}
    \item{Usage of out-of-order delivery mode whenever possible, or limitation on the usage of \gls{mc} in \glspl{ntn} to applications that support such mode}
    \item{Increase the buffer size of \gls{ue}}
    \item{Usage of high timer value to discard outdated data}
    \item{Joint coordination of the transmissions through time compensation}
    \item{Path optimization}
\end{itemize} \\ \hline

Mobility & \gls{ntn} nodes move at a high speed, resulting in frequent \gls{sn} handovers, modifications, and releases. & \begin{itemize}
    \item{Predictive schemes for \gls{mc}-related operations using: 1) \glspl{ue}' GNSS capabilities; and/or 2) satellite ephemeris data}
\end{itemize} \\ \hline

\gls{sn} addition & The standard methods for \gls{sn} addition (based on signal strength) may not be sufficient due to the \gls{ntn} environment. Also, \gls{sn} addition is time consuming. & \begin{itemize}
    \item{\gls{sn} addition based on: 1) location; or 2) elevation angle}
    \item{Optimized \gls{sn} addition without \gls{ra}} 
\end{itemize} \\ \hline

Traffic steering & Traffic steering between the nodes can be difficult due to the long distances. & \begin{itemize}
    \item{Traffic steering schemes that require the minimal amount of control signaling}
    \item{By network configuration, the disablement of \gls{mc} involving nodes that are too far apart} \end{itemize} \\ \hline

Resource allocation & \gls{mn}/\gls{sn} resources used must be coordinated. & \begin{itemize}
    \item{Providing the secondary connection users resources only if there are left from the primary connection users}
    \item{Piggybacking the user’s QoS requirements in the \gls{sn} addition message and taking them into account in resource scheduling on the \gls{sn} side}
    \item{Treating users equally regardless of the connection types}
    \item{Joint optimization of the resource allocation between the \gls{mn} and \gls{sn} via Xn signaling}
    \item{Using beam hopping for increased flexibility in resource allocation}
    \item{In the case of PD, dynamic detection of the need for PD, for example, based on \gls{harq} feedback}
    \end{itemize} \\ \hline

\hline
Power consumption & Power consumption, particularly by \gls{ue} in \glspl{ntn}, poses a significant challenge. & \begin{itemize}
    \item{Usage of the best link at any given time in the uplink direction}
    \item{Utilization of efficient scheduling strategies \cite{10008521} which may include \gls{ai}/\gls{ml}}
\end{itemize} \\ \hline
\end{tabularx}
\end{table*}

\subsection*{Different Types of Nodes Involved in Multi-Connectivity}

The two main satellite payload architectures are transparent and regenerative. Satellites with transparent payloads act as RF repeaters, while the \gls{gnb} (i.e., the \gls{bs} providing 5G access) is on the ground. The \gls{gnb} sends signals through a gateway and the satellite performs frequency conversion and power amplification, and relays the signals. This architecture is considered in \gls{3gpp} \mbox{Rel-17} and \mbox{Rel-18}. In contrast, regenerative payload satellites may have (some of) the \gls{bs}, that is, \gls{gnb}, functionalities on board.

\Gls{mc} in \glspl{ntn} can include only \gls{ntn} nodes or a combination of \gls{tn} and \gls{ntn} nodes. In the former case, the nodes could be transparent or regenerative payload \gls{ntn} nodes or a mix of both. Table II outlines various \gls{mc} scenarios involving \gls{ntn} nodes. Scenarios that involve regenerative payload satellites assume the presence of \gls{pdcp} on board the satellite, otherwise, these cases would be reduced to cases where transparent payload (\gls{pdcp} on the ground) was considered.

Careful consideration must be given when selecting the combination of nodes for \gls{mc} in \glspl{ntn} with respect to the possible need for: 1) increased buffering; 2) formation of \glspl{isl}; and 3) increased path length for the data transmission. For example, regenerative payload \gls{ntn} as a \gls{mn} and transparent payload \gls{ntn} as a \gls{sn} may not be feasible because the user's downlink data directed from the \gls{mn} to the \gls{sn} must first travel to the satellite (\gls{mn}), then to the node on the ground, and then through the transparent payload satellite to the user.

\begin{table*}[hbt!]
\caption{Different \gls{mc} Scenarios Involving \gls{ntn} Nodes Detailed.}
\label{table2}
\begin{tabularx}{\linewidth}{|l|L|L|}
\hline
\textbf{Master node} & \textbf{Secondary node} & \textbf{Considerations} \\ \hline
Transparent payload \gls{ntn} node & Transparent payload \gls{ntn} node & \parnote[1]{Both \glspl{bs} reside on the ground, that is, the Xn interface can be fiber.} \\ \hline
Transparent payload \gls{ntn} node & Regenerative payload \gls{ntn} node & \parnote[2]{Xn must be between the \gls{bs} on the ground and the satellite.} \\ \hline
Transparent payload \gls{ntn} node & Terrestrial node & \footnote[1]{} \\ \hline
Regenerative payload \gls{ntn} node & Transparent payload \gls{ntn} node & This setting could be infeasible because the data directed from the \gls{mn} to the \gls{sn} needs to travel first to the satellite (\gls{mn}), then to the \gls{bs} on the ground, and then to the user through the transparent payload satellite. \newline \footnote[2]{} \\ \hline
Regenerative payload \gls{ntn} node & Regenerative payload \gls{ntn} node & Xn is between the \glspl{bs} onboard the \gls{ntn} nodes, that is, ISL needs to be formed. \\ \hline
Regenerative payload \gls{ntn} node & Terrestrial \gls{bs} & \footnote[2]{} \\ \hline
Terrestrial \gls{bs} & Transparent payload \gls{ntn} node & The terrestrial \gls{bs} may need to consider the extra delay caused by the \gls{ntn} node by buffering some of the data to mitigate the delay spread in some applications where packets must be received sequentially by the user. \newline \footnote[1]{} \\ \hline
Terrestrial node & Regenerative payload \gls{ntn} node & \footnote[2]{} \\ \hline
\end{tabularx}
\parnotes
\end{table*}

\vspace*{1.5\baselineskip}

\subsection*{Delay Differences}

\Gls{nr} \gls{pdcp} supports two types of delivery. The first one is out-of-order delivery, where packets can be delivered to upper layers as they are received without buffering. The delay differences with \gls{mc} in \glspl{ntn} do not cause increased buffering requirements on \gls{ue}’s \gls{pdcp} when this mode is used. When considering using \gls{mc} in \glspl{ntn}, this mode should be used whenever possible. However, this requires the application to support out-of-order packet reception. File transfer or non-live video streaming are examples of such applications. The second delivery mode is in-order delivery, in which \gls{pdcp} packets are buffered at the recipient's \Gls{pdcp} buffer, reordered, and delivered in order. This mode should be used for applications such as voice or live video streaming.

For \glspl{tn}, \gls{ts}~38.331 defines the length of the \gls{pdcp} reordering window, which can vary from 0~ms (out-of-order delivery) to 3~s. Additionally, the window can be set to infinity, which corresponds to forced in-order delivery. These window lengths can cover even the largest delay differences when using \gls{mc} in \glspl{ntn} between heterogeneous nodes, that is, these values can be reused from the \gls{tn} specifications. Since the larger distance between \gls{mn} and \gls{sn} in the \gls{ntn} \gls{mc} case results in larger transmission delay differences than in the \gls{tn} \gls{mc} case, a relatively higher reordering window length should typically be used. This may also require increasing the \gls{pdcp} buffer size of the \gls{ue}.

An alternative option to increasing the \gls{ue}’s \gls{pdcp} buffer size is to coordinate \gls{mn}/\gls{sn} transmissions. This can be accomplished by time alignment or by path optimization, which is the process of choosing the best path for communication based on factors such as congestion levels and propagation delays. However, both methods require signaling between nodes.

\vspace*{\fill}

\subsection*{Mobility}

The beam deployment (shown in Fig.~\ref{beams}) affects the density for \gls{sn} handovers, modifications, and releases. In a quasi-earth fixed beam deployment, the beam pattern is fixed to a target on the ground. In an earth-moving beam deployment, the beams of the satellites are directed in a fixed direction, causing the beams to move on the ground as the satellite moves. In a quasi-earth fixed beam deployment, the \gls{sn} handovers, modifications, and releases may be less frequent than in an earth-moving beam deployment because the \gls{ue} typically stays in the service area longer in a quasi-earth fixed beam deployment.

However, with either beam deployment \gls{sn} handovers, modifications and releases can be highly time consuming, frequent, and caused by satellite movement. In contrast, in \glspl{tn}, \gls{ue} mobility is the primary cause of such events. Indeed, in \glspl{tn} the base stations are static.

Predictive schemes for \gls{sn} handovers/modifications/releases can be used in \glspl{ntn} to anticipate the movement of users/satellites and proactively initiate the necessary \gls{mc}-related operations. For user movement, this can be achieved by exploiting the \glspl{ue}' \gls{gnss} capabilities assumed in \hbox{Rel-17}. For satellite motion, ephemeris data can be used for prediction. This can be done either online by computing the positions of the satellites, or offline by precomputing the positions and using a timer that gives information about the positions of the satellites.

\subsection*{Secondary Node Addition}

The denser deployment of \gls{tn} base stations allows cells to be formed more precisely in the desired locations, thus improving the quality of the links. Due to this setup, cells can be separated with high accuracy and therefore do not interfere with each other even with full frequency reuse. With full frequency reuse, \glspl{ue} can seamlessly collect reference signal strength measurements from adjacent cells without measuring different frequencies.

\begin{figure*}[!t]
\centering
\includegraphics[width=.8\linewidth, frame]{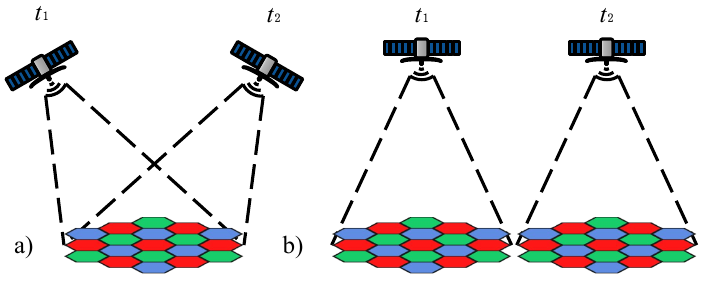}
\caption{a) Quasi-earth fixed and b) earth-moving beam deployments. Frequency reuse 3 is considered where adjacent beams operate at different frequencies corresponding to the different colors to mitigate inter-cell interference. The figure shows the difference in the beam deployments as the satellite moves in space from its initial position between times $t_1$ and  $t_2$.}
\label{beams}
\end{figure*}

In the \gls{ntn} environment, frequency reuse is typically required due to the large diameter of the beams and the relatively small differences in signal strength for cell edge and nadir \glspl{ue}. Frequency reuse is used to reduce inter-cell interference (see Fig.~\ref{beams}). Frequency reuse implies that \glspl{ue} are required to apply measurement gaps to measure the frequencies of adjacent beams. During the measurement gaps, the \gls{ue} may have to stop receiving and/or transmitting data. Therefore, the use of measurement gaps should be carefully designed so that they do not unduly affect the user experience due to paused data reception/transmission. Further, to enable \gls{mc} for a user, the user must be under the coverage of multiple beams. This can be achieved through constellation design and/or directional antennas.

However, even when measurement gaps are carefully designed, the problem of relatively small differences in signal strength between users in a beam remains. Thus, unlike \gls{mc} in \glspl{tn}, where reference signal strength is typically used as a criterion for \gls{sn} addition, other criteria for \gls{sn} addition should be considered when using \gls{mc} in \glspl{ntn}. These criteria might include location or elevation angle.

In addition, the \gls{sn} addition process itself could be optimized. The process in \glspl{tn} is specified in \gls{ts} 37.340. The \gls{sn} addition is initiated by the \gls{mn}, which sends a \gls{sn} addition request to the candidate \gls{sn}. The \gls{sn} then responds with an acknowledgment or rejection. If the request is acknowledged, the \gls{mn} sends an RRC reconfiguration required message to the \gls{ue}. The \gls{ue} performs the required configurations and responds to the \gls{mn}. The \gls{mn} then indicates to the \gls{sn} that the \gls{sn} reconfiguration is complete. Non-contention-based \gls{ra} is used by the \gls{ue} to connect to the \gls{sn}. \Gls{ra} may occur after the \gls{ue} receives the RRC reconfiguration message. The \gls{ra} response contains timing advance and scheduling grant for the \gls{ue} to use for the uplink transmission configuration. In the \gls{sn} addition, the \gls{ra} process takes at least twice the round-trip time, which is a significant time in the \gls{ntn} environment compared to the \gls{tn} environment. Methods to perform handovers without \gls{ra} have already been proposed \cite{7784883}. Similarly, methods to perform \gls{sn} addition without \gls{ra} could be investigated.

\subsection*{Traffic Steering}

Traffic steering refers to steering the user’s traffic from \gls{mn} to \gls{sn}. In \gls{tn} \gls{mc}, it can be done smoothly in a coordinated manner because the \gls{mn} and \gls{sn} are usually located relatively close to each other, allowing for a fast exchange of information. This allows for flexible traffic control strategies that require a lot of control signaling. In \glspl{ntn}, the \glspl{bs} involved in \gls{mc} could even be located on different continents. This imposes limits on the traffic control strategies chosen. Indeed, the same amount of control signaling may not be feasible between the nodes as in the \gls{tn} environment.

This problem can be overcome by using traffic steering schemes that include a minimum amount of control signaling. One option is to use a static split, where the \gls{mn} steers a fixed proportion, say 40\% (the exact amount depends on the use case), of the data for the \gls{sn} to send to the \gls{ue}. By using flow control messages, this split could be made semi-static, making it more flexible, if a bit more complex. Now, the split would otherwise be static, but the \gls{sn} could send flow control messages to the \gls{mn} if the split is too high/low for the \gls{sn}, and a new split ratio would be decided in cooperation between the \gls{mn} and \gls{sn}.

An alternative option to mitigate the traffic steering problem is to use strategic network configuration by disabling \gls{mc} between \glspl{bs} that are not spatially close or sufficiently adjacent in the network infrastructure. This approach allows network administrators to exercise deliberate control by ensuring that \gls{mc} is prohibited between \glspl{bs} whose proximity threshold is not met.

\subsection*{Resource Allocation}

Since secondary connections to a \gls{bs} may affect the existing users, a resource allocation scheme based on connection type, that is, whether the \gls{ue} is a primary connection or a secondary connection \gls{ue}, could be used. Here, a primary \gls{ue} refers to a \gls{ue} that has the \gls{bs} as its \gls{mn}, and a secondary \gls{ue} refers to a \gls{ue} that has the \gls{bs} as its \gls{sn}. Prioritizing \glspl{ue} with primary connections in resource allocation can be used to avoid impacting existing users. This means that leftover capacity from the primary connection \glspl{ue} will be distributed to the secondary connection \glspl{ue}, rather than affecting the \glspl{ue} already being served by the \gls{bs}. Another solution would be to treat all connections equally and not distinguish between primary and secondary \glspl{ue} connections. However, the secondary connection users typically have worse channel conditions than the primary connection users, which means that the secondary connection users require more resources.

An example of a more complex resource allocation scheme includes one where the quality of service (QoS) requirements of the \glspl{ue} can be taken into account. One way to accomplish this would be to piggyback this information into the \gls{sn} addition request message. Further, if the \gls{sn} and \gls{mn} can exchange information about their resource scheduling strategies, the resource allocation at each node can be jointly optimized via Xn signaling.
In addition, beam hopping can be used to switch beams between different coverage areas. This allows for flexible provisioning of resources according to demand. MC could be used in conjunction with beam hopping, and by intelligently designing the beam hopping patterns, more flexibility in resource allocation could be introduced.

\Gls{mc} can also be used to enhance reliability through \gls{pd} \cite{8884124}. In \gls{pd}, a packet is duplicated and transmitted over different links to increase the probability that the transmission over at least one of the links is successful. While \gls{pd} enhances reliability it also severely increases resource consumption. Instead of using blind \gls{pd}, in which all of the packets are duplicated for a given user, dynamic schemes for detecting the need for \gls{pd} should be utilized. One of the suggested schemes includes \gls{pd} activation based on \gls{harq} feedback \cite{10195478}.

\subsection*{Power Consumption}

Power consumption is a significant issue, especially on the \gls{ue} side in \glspl{ntn}, especially when \gls{mc} is used in the uplink direction with simultaneous transmission to \gls{mn} and \gls{sn}. However, \gls{mc} can be used in the uplink by selecting the best link in terms of SINR at any given time. This could alleviate the power consumption problem on the \gls{ue} side, since fewer resources are needed when the quality of the link is better. In this approach, \gls{mc} is used to enable a backup link, which also contributes to service continuity if a link fails. In addition, a likely Rel-19 feature called High Power UE (HPUE), which allows a UE to transmit at higher power, helps to enable simultaneous transmission to MN and SN.

The use of efficient scheduling strategies \cite{10008521} can lead to efficient resource utilization and an increase in spectral efficiency. Indeed, the use of \gls{mc} can lead to a greater number of possible paths through which the data can be routed. This results in a complex path selection process that requires intelligent scheduling strategies. \Gls{ai}/\gls{ml} can be used to solve such complex problems, for example, by learning online from the environment or by learning offline from historical data. It should be noted that \gls{ai}/\gls{ml} is a tool that can also be used to help solve the problems listed in the previous subsections.

\section*{Conclusion}

The convergence of \glspl{ntn} and \glspl{tn} into a unified network that is indistinguishable to the user is underway. To efficiently utilize the available resources, the use of disruptive technologies is required. One of these is \Gls{mc} in \glspl{ntn}, where a user can be connected to multiple \glspl{bs}, which can include \gls{ntn} and \gls{tn} nodes.

Challenges that need to be overcome to achieve \gls{mc} in \glspl{ntn} have been discussed in this article and solutions have been proposed. The proposed solutions help pave the way for future research that needs to consider the technical details of the solutions. By overcoming the identified challenges, the use of \gls{mc} in \glspl{ntn} can bring significant benefits to various applications. These include, but are not limited to, load balancing, throughput enhancement for cell edge users to enable high throughput applications such as live video streaming, and livestock and environmental monitoring.

While this article primarily examines challenges related to the application of \gls{pdcp} layer \gls{mc} in \glspl{ntn}, it is important to note that some of the proposed solutions may also be relevant to \gls{mc} implementations on other layers. In future work, there is a need for a comprehensive analysis of \gls{mc} in \glspl{ntn} applied on different layers, but also the potential cross-layer applicability of the solutions proposed in this article. Further, the consideration of \gls{mc} in multi-layer \gls{ntn} scenarios should also be included in future research.

\section*{Acknowledgment}
The author thanks Prof. Timo Hämäläinen and Dr. Jani Puttonen, his Ph.D. supervisors, for their valuable feedback on the article.

\vspace{6pt}
\bibliography{references} 
\bibliographystyle{IEEEtran}


\section*{Biographies}

\noindent \textsc{Mikko Majamaa} [STM'23] (mikko.majamaa@magister.fi) works as a senior researcher at Magister Solutions. He is also a Ph.D. candidate at the University of Jyväskylä, Finland. His research interests include multi-connectivity, radio resource management, and the use of machine learning in 5G and beyond non-terrestrial networks.

\end{document}